\documentclass[aps,twocolumn,superscriptaddress,a4paper,floatfix]{revtex4-1}
\usepackage[latin1]{inputenc}
\usepackage[pdftex]{graphicx}
\usepackage{amsmath,amssymb}
\usepackage{hyperref}

\newcommand \beq{\begin{eqnarray}}
\newcommand \eeq{\end{eqnarray}}

\begin{document}

\title{Role of Nambu-Goldstone modes  in the fermionic superfluid point contact}

\author{Shun Uchino}
\affiliation{Waseda Institute for Advanced Study, Waseda University, Shinjuku, Tokyo 169-0051, Japan}


\begin{abstract}
In fermionic superfluids that are charge neutral, Nambu-Goldstone (NG) modes also known as Anderson-Bogoliubov modes 
emerge as a result of spontaneous symmetry breaking.
Here, we discuss DC transport properties of  such NG modes through a quantum point contact.
We show that
contrary to a naive view that enhancement of the phase stiffness may suppress transport of the NG modes, there must be an
anomalous contribution that survives at low temperature.
This contribution originates from the conversion process between the condensate and NG mode.
We find that  within the BCS regime the anomalous contribution is enhanced with increasing channel transmittance and attractive interaction, and
leads to a temperature-dependent Lorenz number and absence of the bunching effect in current noise.

\end{abstract}

\maketitle

\section{Introduction}

In mesoscopic transport phenomena through small constrictions, 
quantum mechanical effects are known to be directly reflected in transport coefficients.
One of the best known is the Landauer formula in which the two-terminal conductance of normal metals is quantized~\cite{landauer}.
This quantization originates from disappearance of relevant length scales in constriction due to the ballistic condition and therefore
is clearly seen in a quantum point contact where a constriction has a short one-dimensional structure~\cite{datta1997,nazarov2009}.

In addition to shapes of constriction, states of matter in reservoirs play an important role in mesoscopic transport.
The prototype example is a superconducting point contact where reservoirs consist of superconductors~\cite{PhysRevLett.73.2611,PhysRevLett.78.3535}.
In this case, it is known that the direct current does not obey Ohm's law~\cite{PhysRevLett.73.2611,PhysRevLett.78.3535}.
The key ingredient there is multiple Andreev reflections~\cite{klapwijk} where  quasiparticles repeat Andreev reflections at the boundaries between
superconductor and contact. As a result, the current-bias characteristics become highly nonlinear~\cite{PhysRevLett.75.1831}.

In contrast, each constituent particle in detail is expected to be irrelevant in mesoscopic transport.
For instance, when electrons are replaced by other fermions e.g. neutral atoms such as $^6$Li and $^{40}$K,
essentially the same phenomena are observed as long as similar states of matter are prepared.
This type of universality can nowadays be confirmed with ultracold atomic gases.
Indeed, a two-terminal transport setup with a quantum point contact 
has been realized in experiments of ultracold Fermi gases~\cite{krinner2017}, which 
observed the conductance quantization~\cite{krinner2015} and nonlinear current-bias characteristics~\cite{husmann2015}.

It must be noted, however, that the presence or absence of charge may cause a  difference in transport between electron and atomic systems.
Especially, this difference may qualitatively be important for systems with Bose-Einstein condensation of Cooper pairs where
Nambu-Goldstone (NG) modes emerge due to spontaneous symmetry breaking~\cite{nagaosa2013}.
In the case of electrons (superconductors), an effect of the Coulomb interaction is inevitable, which turns
the NG modes into the gapped plasma modes~\cite{PhysRev.112.1900}.
In the case of atoms (superfluids), on the other hand, these NG modes also known as Anderson-Bogoliubov modes 
remain gapless~\cite{PhysRev.112.1900,bogoliubov}, and the NG modes may play an important role in low-energy transport.
At the same time, as the NG modes are a non-superfluid component,  there is a view that  the effects of such modes are
negligible at low temperature.
Since  the NG mode in mesoscopic transport have yet to be incorporated in an explicit manner,
 it is not clear whether it is reasonable to neglect the effect of the gapless mode in experiments of ultracold atomic gases.
 
\begin{figure}[t]
\begin{center}
\includegraphics[width=7cm]{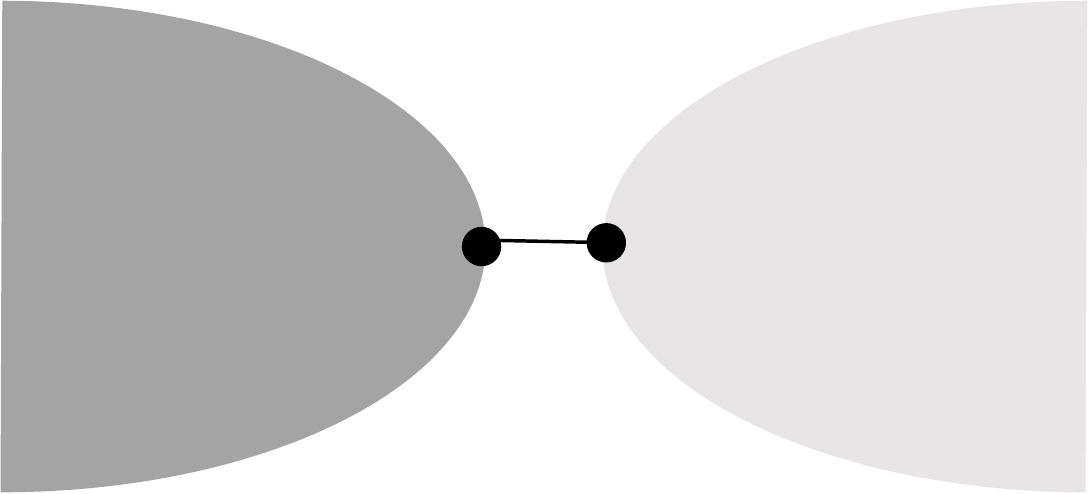}
\caption{Schematic illustration of the fermionic superfluid point contact, where two macroscopic reservoirs are
fulfilled by $s$-wave Fermi superfluids.
We analyze a DC current including transport of the Nambu-Goldstone modes 
in terms of the tunneling Hamiltonian, provided that
the constriction length is shorter than the superfluid coherence length.}
\label{fig} 
\end{center}
\end{figure}

In this paper, we analyze direct currents pertinent to transport of the NG modes in the fermionic superfluid point contact, as illustrated in Fig.~\ref{fig}.
On one hand, we demonstrate that the naive view mentioned above is partially correct in that the exchange process
of the NG modes between reservoirs is indeed suppressed at low temperature.
On the other hand, we uncover an anomalous process such that condensate elements are converted into
the NG modes and vice versa.
What is remarkable is  that this process survives even at absolute zero.
It is discussed that within the BCS regime the anomalous contribution is enhanced with increasing channel transmittance and attractive interaction of fermions, and
is absent in the heat current, which causes breakdown of the Wiedemann-Franz law.

This paper is organized as follows.
Section II discusses an effective action of the NG modes in fermionic superfluids in terms of functional integrals.
In Sec. III, we introduce the tunneling Hamiltonian approach to discuss transport of the NG modes
in the superfluid point contact.
In Sec. IV, the pair current expression including effects of the NG modes is obtained by using the analyses in Sec. II and  Sec. III.
Section V discusses several topics related to the results of Sec IV.
Section VI summarizes this paper and the calculation of correlation functions is shown in detail in Appendix A.

\section{Effective field theory of Anderson-Bogoliubov modes}
Here, we sketch how the effective action of the Anderson-Bogoliubov modes  in the bulk (each reservoir) is obtained 
from fermionic superfluids.
To this end, we consider the following bulk Hamiltonian (we use units $\hbar=k_B=1$):
\beq
H_0&&=\int d^3x{\cal H}_0,\\
{\cal H}_0&&=\sum_{\sigma=\uparrow,\downarrow}\psi^{\dagger}_{\sigma}\Big(-\frac{\nabla^2}{2M}-\mu\Big)\psi_{\sigma}+g\psi^{\dagger}_{\uparrow}\psi^{\dagger}_{\downarrow}
\psi_{\downarrow}\psi_{\uparrow},
\eeq
where $\psi_{\sigma}$ is the fermionic field with spin $\sigma$,
$M$ is the mass of fermions, $\mu$ is the chemical potential. In addition, the coupling constant
$g$ is assumed to be negative to ensure an attractive interaction between fermions.
We note that the above Hamiltonian is nothing but the celebrated BCS Hamiltonian and is also known to describe
two-component Fermi gases interacting via broad Feshbach resonances~\cite{RevModPhys.80.885}.

We focus on the low-temperature regime where $\uparrow$ and $\downarrow$ fermions form Cooper pairs and 
Bose-Einstein condensation of Cooper pairs occurs. When the attractive interaction between fermions is weak (BCS regime),
such condensation can be captured by the mean-field theory~\cite{zwerger2011}. 
By adopting this theory, the original Hamiltonian reduces to 
\beq
{\cal H}_0\to\sum_{\sigma}\psi^{\dagger}_{\sigma}\Big(-\frac{\nabla^2}{2M}-\mu\Big)\psi_{\sigma}-\Delta\psi^{\dagger}_{\uparrow}\psi^{\dagger}_{\downarrow}
-\bar{\Delta}\psi_{\downarrow}\psi_{\uparrow}.
\label{eq:mfh}
\eeq
Here, we introduce the gap parameter,
\beq
\Delta =-g\langle \psi_{\downarrow}\psi_{\uparrow} \rangle,
\eeq
which is assumed to be a constant in spacetime.
Since the Hamiltonian  above is quadratic in $\psi$, we can obtain a quasiparticle excitation by diagonalizing it.
The resultant quasiparticle is called Bogoliubov mode that is fermionic and has the energy gap $2|\Delta|$.

In addition to the fermionic quasiparticle excitation, there are bosonic collective excitations in fermionic superfluids. 
The NG mode is then the dominant excitation at low energy, since it
is gapless due to the NG theorem~\cite{nagaosa2013}.
To see this gapless mode, we consider fluctuations from the mean field in the gap function. 
In general, the gap parameter is complex and therefore
there are two directions of fluctuations; amplitude fluctuation and
phase fluctuation, which are related to Higgs mode and NG mode, respectively.
Since the former excitation has a gap $2|\Delta|$~\cite{pekker2015}, 
we can focus on the phase fluctuation in  the  low-frequency regime 
$|\omega|< |\Delta|$.
In this case, the gap parameter with the phase fluctuation is expressed as follows:
\beq
\Delta(x)=|\Delta|e^{2i\phi(x)},
\eeq
where $\phi$ is a real bosonic field describing the phase fluctuation.
To obtain an effective theory of the NG mode, it is convenient to adopt the functional integral formalism.
By integrating out the fermionic fields in the mean-field Hamiltonian~\eqref{eq:mfh}, 
the partition function can be expressed in terms of $\phi$ such that
\beq
Z= \int {\cal D}\phi e^{-S_{\text{eff}}(\phi)}.
\eeq
Here, $S_{\text{eff}}$ is the effective action of the NG mode, and is in general  a complicated function in $\phi$.
By performing the leading-order analysis of the gradient approximation in $\phi$, however, $S_{\text{eff}}$
becomes the following quadratic form~\cite{nagaosa2013}:
\beq
S_{\text{eff}}\approx \frac{\rho_s}{2}\int_0^{\beta} d\tau\int d^3x\Big(\frac{1}{v^2}(\partial_{\tau}\phi)^2+(\nabla \phi)^2\Big),
\label{eq:effective-action}
\eeq
where $v$ is the speed of sound and  $\rho_s$ is the superfluid density.
At absolute zero, they are explicitly determined as
\beq
&&v=\frac{v_F}{\sqrt{3}},\\
&&\rho_s=\frac{n}{M},
\eeq
with the Fermi velocity $v_F$ and number density $n$.

The procedure introduced above is called bosonization in  the way that the effective theory of the bosonic collective mode is obtained 
from the fermionic action.
In our derivation, the mean-field theory and gradient approximation are explicitly employed.
Regardless of strength of the attractive interaction, however, the superfluid phase with
 spontaneous breaking of U(1) symmetry is known to emerge in the low-temperature regime of the spin-balanced mixture~\cite{RevModPhys.80.885}.
This means that the NG modes with the linear gapless dispersion are always present in the superfluid phase~\cite{watanabe}.
Indeed, as far as the low-energy regime is concerned,  the form of the effective action \eqref{eq:effective-action} is universal for the U(1) symmetry breaking phase~\cite{sachdev}.
Then, effects on the interaction and temperature are reflected as renormalization of $\rho_s$ and $v$.

\section{Tunneling Hamiltonian approach}
Now that  the effective action of the NG mode in the bulk superfluid is obtained,
we wish to discuss how this gapless mode affects two-terminal point contact transport, where two macroscopic reservoirs (L and R) are connected through
a short one-dimensional wire.
In particular, we focus on a regime where the length of the one-dimensional wire is shorter than the superfluid coherence length $ v_F/(\pi |\Delta|)$.
The constriction in detail is then known to be irrelevant~\cite{RevModPhys.51.101} 
and transport of such a system can be discussed with the following 
tunneling Hamiltonian~\cite{PhysRevB.54.7366,PhysRevB.84.155414,PhysRevLett.118.105303}:
\beq
H=H_0+H_T, \\
H_{0}=H_L+H_R
\eeq
where $H_{L(R)}$ is the grand Hamiltonian of the left (right) reservoir, and
\beq
H_T=t\sum_{\sigma}\psi^{\dagger}_{\sigma, R}(\mathbf{0}) \psi_{\sigma,L}(\mathbf{0})+h.c.
\eeq
is the tunneling term with the tunneling amplitude $t$.
Since in this model, a particle exchange occurs at the single point $\mathbf{x=0}$,
it follows that the currents are expressed by the fields at $\mathbf{x=0}$.
Thus, in what follows, we omit the spatial index in fields  for brevity.
Based on this Hamiltonian, the particle current operator is calculated as
\beq
I&&=-\dot{N}_L=i[N_L,H_T]\nonumber\\
&&=-it\sum_{\sigma}\psi^{\dagger}_{\sigma,R}\psi_{\sigma,L}+h.c.
\eeq
We note that the tunneling term above represents the single-particle tunneling between the reservoirs.
Therefore, the direct current calculated with the mean-field Hamiltonian~\eqref{eq:mfh}
turns out to be associated with quasiparticle processes including multiple Andreev reflections~\cite{PhysRevB.54.7366}.

In addition to the quasiparticle tunneling, the tunneling Hamiltonian allows us to discuss
the pair tunneling process related to transport of the NG modes.
The key point is that in the tunneling Hamiltonian approach, the tunneling term is treated as perturbation and therefore
the pair tunneling process is generated as a higher-order tunneling effect~\cite{giamarchi}.
To see this in an explicit manner, we temporarily consider the zero-bias situation, where a contribution from left to right is balanced with
one from right to left. By using the imaginary time formalism, a contribution related to the NG modes is extracted as
\beq
\langle I_{\uparrow}\rangle&&=-it\langle\sum_{n=0}^{\infty}\frac{(-1)^{n+1}}{(n+1)!}\int_0^{\beta}d\tau_1\cdots\int_0^{\beta}d\tau_{n+1}\nonumber\\
&&\times T_{\tau}[ \psi^{\dagger}_{\uparrow,R}\psi_{\uparrow,L} H_T(\tau_1)\cdots H_{T}(\tau_{n+1})]
\rangle_0+h.c.\nonumber\\
&&\to \frac{i\alpha t^2\Delta_L\Delta_R}{g^2}\langle e^{-2i\phi_R}e^{2i\phi_L} \rangle
+h.c.,
\label{eq:trick}
\eeq
where $\langle\cdots\rangle_0$ means the average without the tunneling term, and
in the last line of the equation, we extract the pair tunneling process and
introduce the short time scale $\alpha$ for the extraction.
Since the similar result is obtained for $\langle I_{\downarrow}\rangle$,
the pair current operator related to transport of the NG  modes is obtained as
\beq
I_{p}=-it_p e^{-2i\phi_R}e^{2i\phi_L}+h.c.,
\label{eq:pair-current}
\eeq
with $t_p=-2\alpha t^2\Delta_L\Delta_R/g^2$.
In addition,  $H_T$ generates the pair tunneling term, showing up even order in $t$.
Indeed, by using the similar trick used in Eq.~\eqref{eq:trick}, 
the imaginary time evolution operator of even order is transformed into
\beq
\sum_{n=2,4,\cdots}\int_0^{\beta}d\tau_1\cdots\int_0^{\tau_{n-1}} d\tau_nH_T(\tau_1)\cdots H_T(\tau_n)\nonumber\\
\to\sum_{n=1}^{\infty}\int_0^{\beta}d\tau_1\cdots\int_0^{\tau_{n-1}}d\tau_n H_{p}(\tau_1)\cdots H_{p}(\tau_n),
\eeq
where
\beq
H_{p}=t_pe^{-2i\phi_R}e^{2i\phi_L}+h.c.,
\eeq
is the pair tunneling term. Thus, the discussions above implies that the pair tunneling contribution of the mass current
is obtained as the average of Eq.~\eqref{eq:pair-current} under the perturbation term $H_p$.

\section{Pair current expression}
We now discuss a pair current expression in the presence of a chemical potential bias, $\Delta\mu=\mu_L-\mu_R$.
In order to avoid a contribution of the Higgs modes, we postulate the condition $\Delta\mu, T <\Delta$.

We  note that $\Delta\mu$ is the bias for each fermion.
Therefore, when it comes to the pair tunneling, the bias between the reservoirs must be regarded as $2\Delta\mu$.
We can also understand this result in terms of the gauge transformation technique conventionally utilized in the tunneling Hamiltonian 
approach, where the combination $\psi^{\dagger}_{\sigma,R(L)}(\tau)\psi_{\sigma,L(R)}(\tau)$ yields the factor $e^{-(+) i\Delta\mu\tau}$~\cite{mahan2013}.
Then, since the pair tunneling process is related to the combination $\psi^{\dagger}_{\uparrow,R(L)}(\tau)\psi^{\dagger}_{\downarrow,R(L)}(\tau) 
\psi_{\downarrow,L(R)}(\tau)\psi_{\uparrow,L(R)}(\tau)$, such a term gives rise to the factor  $e^{-(+) 2i\Delta\mu\tau}$, meaning that
the chemical potential bias on the pair tunneling is $2\Delta\mu$~\cite{PhysRevLett.118.105303}.

To obtain a pair current expression, we  also note that
the average of the current at real time $\tau$ can be expressed as
\beq
\langle I_{p}(\tau) \rangle= 2\text{Re}\Big[t_pG^<_{LR}(\tau,\tau)  \Big],
\eeq
with  lesser Green's function 
\beq
G^<_{LR}(\tau,\tau')=-i\langle e^{-2i\phi_R(\tau')}e^{2i\phi_L(\tau)} \rangle.
\eeq
By using the expression above, we turn to calculate the current expression including arbitrary order in $t_p$.
To this end, it is important to recall the following properties of uncoupled retarded and advanced Green's functions under the gaussian action~\cite{giamarchi}:
\beq
g^R_{L(R)}(\tau)=-i\theta(\tau)\langle [e^{Ai\phi_{L(R)}(\tau)}, e^{Bi\phi_{L(R)}(0)}] \rangle_0=0,\label{eq:g1}\\
g^A_{L(R)}(\tau)=i\theta(-\tau)\langle [e^{Ai\phi_{L(R)}(\tau)}, e^{Bi\phi_{L(R)}(0)}] \rangle_0=0,\label{eq:g2}
\eeq
unless $A+B=0$. 
These properties on the average without $H_p$ forbid emergence of the anomalous average contribution appearing in the quasiparticle contribution 
in superconducting systems~\cite{PhysRevB.54.7366}, and  renders the current calculation simple.

In order to obtain the current expression including an arbitrary order in $t_p$, we consider the Dyson equation in the real-time formalism~\cite{rammer2007}.
By using the so-called Langreth rules~\cite{rammer2007}, the Dyson equation of $G^<$ is obtained as
\beq
G^<=(1+G^RV)\circ g^<\circ (1+VG^A),
\eeq
where $\circ$ denotes integration over the internal time variable from minus infinity to plus infinity, $V$ is one-particle potential, 
$G^{R(A)}$ is exact retarded (advanced) Green's function including the effect of $V$, and $g^<$ is uncoupled lesser Green's function.
By applying the Dyson equation above for the tunneling Hamiltonian where $V$ is replaced by the tunneling amplitude, we obtain
\beq
t_pG^<_{LR}=g^R_{L}\circ \bar{T}^R_p\circ g^<_{R}\circ T^A_p+T^R_p\circ g^<_{L}\circ \bar{T}^A_p\circ g^A_{R}.
\eeq
Here, we introduce the renormalized tunneling amplitudes
\beq
T_p^{R,A}=t_p+t_pg^{R,A}_{L}\circ \bar{t}_pg^{R,A}_{R}\circ T_p^{R,A},\\
\bar{T}_p^{R,A}=\bar{t}_p+\bar{t}_pg^{R,A}_{R}\circ t_pg^{R,A}_{L}\circ \bar{T}_p^{R,A}.
\eeq
By  performing the Fourier transformation,
the renormalized tunneling amplitudes can be solved as
\beq
T_p^{R,A}(\omega)=\frac{t_p}{1-|t_p|^2g^{R,A}_{L}(\omega-2\Delta\mu)g^{R,A}_{R}(\omega)},\\
\bar{T}_p^{R,A}(\omega)=\frac{\bar{t}_p}{1-|t_p|^2g^{R,A}_{L}(\omega-2\Delta\mu)g^{R,A}_{R}(\omega)}.
\eeq
To obtain above, we use the fact that tunneling is an energy conserving process and origins of frequencies between the reservoirs  are different  by $2\Delta\mu$.
Thus, the DC pair current is obtained as
\beq
&&\langle I_{p}(\tau) \rangle=-\int_{-\infty}^{\infty}\frac{d\omega}{2\pi} \frac{2|t_p|^2}{|1-|t_p|^2g^{R}_{L}(\omega-2\Delta\mu)g^{R}_{R}(\omega)|^2}
\nonumber\\
&&(\text{Im}[g^R_{L}(\omega-2\Delta\mu)] \text{Im}[g^<_{R}(\omega)] +\text{Im}[g^A_{R}(\omega)] \text{Im}[g^<_{L}(\omega-2\Delta\mu)] ).\nonumber\\
\label{eq:general-mass}
\eeq

For the sake of qualitative discussions on the transport properties, we consider the small bias regime where $O(\Delta\mu^2)$ is negligible.
By using the expressions of Green's functions obtained in Appendix A, we reach the following linear current-bias relation:
\beq
\langle I_p\rangle=(G_{\text{an}}+G_{\text{NG}})\Delta\mu.
\label{eq:I-V}
\eeq
Here, we classify the conductance into two contributions, since the lesser Green's function contains a part proportional to a condensate $\delta(\omega)$ and
one proportional to the phonon distribution  $\displaystyle n(\omega)=\frac{1}{e^{\omega/T}-1}$.
The former yields $G_{\text{an}}$ given by
\beq
G_{\text{an}}={\cal T}^2\frac{2M^2v^3}{\pi\rho_s},
\eeq
where
\beq
{\cal T }^2=\frac{4|t_p|^2/(M^2v^4)}{|1-|t_p|^2g^{R}_L(0)g^{R}_R(0)|^2}
\label{eq:transparency}
\eeq
is dimensionless parameter related to the channel transmittance.
This contribution is anomalous in that condensation causes the direct current and does not vanish even at absolute zero.
Physically, $G_{\text{an}}$ is related to the conversion process between the condensate and NG mode, which also appears in transport
of bosonic systems~\cite{PhysRevA.64.033610,uchino}.
On the other hand, the latter originates from  normal tunneling of the NG modes between the reservoirs.
At a low temperature,  $G_{\text{NG}}$ is reduced to the following simple form:
\beq
G_{\text{NG}}&&= \int_{-\infty}^{\infty}\frac{d\omega}{2\pi} \frac{8|t_p|^2\text{Im}[g^R(\omega)]^2
}{|1-|t_p|^2g^{R}_L(\omega)g^{R}_R(\omega)|^2}\Big(-\frac{\partial n(\omega)}{\partial\omega}\Big)\nonumber\\
&&\approx {\cal T}^2\frac{2M^2v^2T^2}{3\pi \rho_s^2},
\eeq
where we use
\beq
-\frac{\partial n(\omega)}{\partial\omega}=\frac{1}{4T\sinh^2(\omega/(2T))}.
\eeq
In contrast to the anomalous contribution, the contribution above explicitly depends on temperature 
in such a way that  it vanishes at zero temperature.
This is due to the fact that the NG modes that are the non-superfluid components must be  suppressed at a low temperature.

\section{discussion}
\begin{figure}[t]
\begin{center}
\includegraphics[width=7cm]{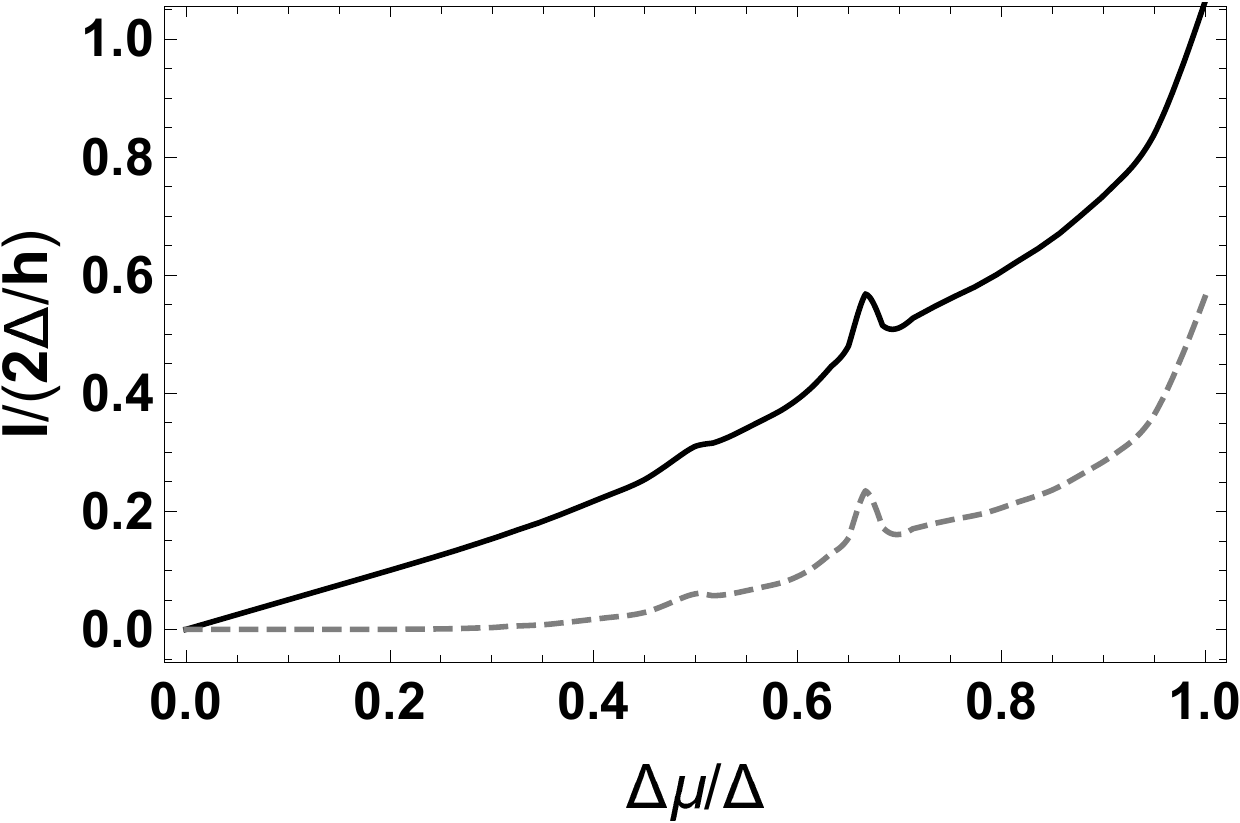}
\caption{Typical behavior of the current-bias characteristics in the fermionic superfluid point contact at $T=0$. The dashed curve represents the quasiparticle current  in which the transmittance  is given as $0.5$~\cite{PhysRevB.54.7366} 
\footnote{In the tunneling Hamiltonian approach, the channel transmittance of each fermion is expressed as $\frac{4t^2/(\pi\rho_q(\mu_F))^2}{(1+t^2/(\pi\rho_q(\mu_F))^2)^2}$
with the single-particle density of states at the Fermi level $\rho_q(\mu_F)$}. 
 On the other hand, the total contribution being the sum of the quasiparticle and pair currents is plotted in the solid curve, where
$G_{\text{an}}=1/h $.}
\label{fig2} 
\end{center}
\end{figure}

Here, we discuss some implications based on the results obtained in the previous sections.

First, it turns out that the contributions related to the NG modes obey Ohm's law at small biases,
in contrast to the quasiparticle contribution, which is essentially nonlinear in $\Delta\mu$.
As can be seen from the definition of $t_p$ and Eq.~\eqref{eq:transparency},
the leading order term in the pair current is proportional to $t^4$.
In order to measure the NG mode contribution, we therefore look at the high transmittance regime beyond the tunneling limit 
proportional to $t^2$. 

As shown in the dashed line of Fig.~\ref{fig2},
in the intermediate transmittance regime, 
the quasiparticle current shows
the subharmonic gap structure at $\Delta\mu/\Delta=2/n$ ($n=1,2,\cdots$)  due to the multiple Andreev reflections
and is still negligible at the small bias.
On the other hand, in the BCS limit at zero temperature,  we obtain $G_{\text{an}}/{\cal T}^2\sim 4\pi^2/(\sqrt{3} h)$.
Since we apply 
the effective field theory approach, which inevitably contains the short range cutoff,
it is difficult to a priori obtain the quantitative value of the conductance.
At the same time, this estimation  implies that the bias-independent conductance of the order of $1/h$ arises from the anomalous contribution.
In this regime at small bias, 
it is thus expected that the mass current is dominated by
the Ohmic signal originating from the NG modes (see the solid line of  Fig.~\ref{fig2}).

Next, we note that $t_p$ depends on the interatomic interaction $g$. By using the gap formula in the BCS theory $\Delta\sim e^{1/(gN(0))}$ with the density of state at the Fermi level $N(0)$, the pair tunneling amplitude is found to behave as $\displaystyle t_p\sim \frac{e^{2/(gN(0))}}{g^2}$.
This  coupling dependence in $t_p$ shows that  when increasing the strength of the coupling constant $|g|$,
the pair tunneling effect is enhanced at least within the BCS regime.  Physically, this means that with increasing $|g|$, the size of the Cooper pairs decreases and the pair tunneling event through the point contact tends to occur. 
In addition,  at the zero temperature limit where $\rho_s=\rho$, the conductance turns out to be inversely proportional to the speed of sound $v$. A recent experiment on the Bragg spectroscopy confirms that $v$ is decreased with increasing strength of the coupling $|g|$~\cite{hoinka2017}. Therefore, as far as the BCS regime is concerned,
 the anomalous contribution is expected to be enhanced with increasing $|g|$.

In what follows, we  address effects of temperature bias and electric charge.

\subsection{Effect of a temperature bias}
We now consider a situation without chemical potential bias but with temperature bias $\Delta T=T_L-T_R$.
As in the case of the chemical potential bias, we focus on the regime $\Delta T,T<\Delta$.

First, we discuss the effect of $\Delta T$ on the mass current.
To this end, we note that the DC pair current expression~\eqref{eq:general-mass} is available even in the presence of $\Delta T$.
This is because in the two-terminal system a temperature is fixed in each reservoir, and 
the calculations of the correlation functions without the tunneling term discussed in Appendix A are available regardless of presence or absence of $\Delta T$.
Then, up to linear order in $\Delta T$, the pair current can be expressed as
\beq
\langle  I_p\rangle =(L_{12,\text{an}}+L_{12,\text{NG}})\Delta T,
\eeq
where $L_{12,\text{an}}$ and $L_{12,\text{NG}}$ are respectively anomalous and normal NG mode contributions similar to Eq.~\eqref{eq:I-V}.
By using the correlation functions obtained in Appendix A, we find 
\beq
L_{12,\text{an}}\propto\int d\omega\omega \delta(\omega)=0.
\eeq
Thus, the anomalous contribution is not induced by $\Delta T$.
We note that this is consistent with the result found in bosonic superfluid point contact~\cite{uchino}.
Similarly, normal contribution of the NG modes at a low temperature is obtained as
\beq
L_{12,\text{NG}}\propto \int d\omega \frac{\omega^3}{\sinh^2(\omega/2T)}\to0,
\eeq
where we use
\beq
\frac{\partial n(\omega)}{\partial T}=\frac{\omega}{4T^2\sinh^2(\omega/(2T))}.
\eeq
Here, we point out that the absence of $L_{12,\text{NG}}$ is a consequence of the quadratic action~\eqref{eq:effective-action},
which predicts $\text{Im}[g^R(\omega)]\propto \omega$.
In general, there must be effects beyond quadratic theory such as interactions between the NG modes
 leading to nonzero value of  $L_{12,\text{NG}}$.
In a low enough temperature where the NG modes are rarely excited,
however, such effects are negligible, and  the Seebeck effect related to the NG modes is absent.

Next, we discuss the heat current induced by $\Delta T$.
In the tunneling Hamiltonian, the heat current operator is calculated as
\beq
I_Q&&=-\dot{H}_L\nonumber\\
&&=t\sum_{\sigma}\psi^{\dagger}_{\sigma,R}\frac{d}{d\tau}\psi_{\sigma.,L}+h.c.,
\eeq
where we use
the Heisenberg equation of motion for $\psi_{\sigma,L}$.
By using the similar trick used in the mass current, 
the heat pair current may be obtained as~\cite{uchino}
\beq
\langle I_{p,Q}(\tau) \rangle=2\lim_{\tau'\to\tau}\text{Re}\Big[it_p\frac{d}{d\tau}G^<_{LR}(\tau,\tau')\Big].
\eeq
In the frequency space, the expectation above is simplified as
\beq
&&\langle I_{p,Q}\rangle=-\int_{-\infty}^{\infty}\frac{d\omega}{2\pi} \frac{2\omega |t_p|^2}{|1-|t_p|^2g^{R}_{L}(\omega)g^{R}_{R}(\omega)|^2}
\nonumber\\
&&\times(\text{Im}[g^R_{L}(\omega)] \text{Im}[g^<_{R}(\omega)] +\text{Im}[g^A_{R}(\omega)] \text{Im}[g^<_{L}(\omega)] ).
\eeq
We now explicitly consider the linear regime in $\Delta T$.
Then, the heat pair current can be expressed as
\beq
\langle I_{p,Q}\rangle=(L_{22,\text{an}}+L_{22,\text{NG}})\Delta T.
\eeq
The anomalous contribution of the heat current is shown to vanish, since 
\beq
L_{22,\text{an}}\propto\int d\omega\omega^2\delta(\omega)=0.
\eeq
On the other hand, the normal contribution of the heat current is given by
\beq
L_{22,\text{NG}}&&=\int_{-\infty}^{\infty}\frac{d\omega}{2\pi} \frac{4 \omega |t_p|^2\text{Im}[g^R(\omega)]^2
}{|1-|t_p|^2g^{R}_L(\omega)g^{R}_R(\omega)|^2}\Big(\frac{\partial n(\omega)}{\partial T}\Big)\nonumber\\
&&\approx{\cal T}^2\frac{4\pi M^2 v^2T^3}{15\rho_s^2}.
\eeq

Discussions above mean that the thermal conductance induced by the NG modes
reduces to $L_{22,\text{NG}}$. Moreover, the corresponding Lorenz number is obtained as
\beq
L_p\approx \frac{L_{22,\text{NG}}}{TG_{\text{an}}}\approx\frac{2\pi^2T^2}{15v\rho_s}.
\eeq
It is notable that the Lorenz number depends on thermodynamic quantities such as temperature, $v$, and $\rho_s$.
Thus, this is quite different from the Wiedemann-Franz law where the Lorenz number is expressed in terms of fundamental constants in physics 
such as the Boltzmann constant.
As in the case of the bosonic superfluid case, the breakdown of the Wiedemann-Franz law is attributed to the presence of 
the anomalous contribution of the mass current that cannot be interpreted as simple quasiparticle tunneling between the reservoirs~\cite{uchino}.
Moreover, the expression above shows that the Lorenz number becomes small at  a low temperature, which is consistent with
the recent experimental observation~\cite{husmann2018}.

\subsection{Current noise}
It is interesting to look at the current noise behavior induced by the pair current~\footnote{While an experimental 
observation of such a correlation function in the cold atomic setup 
has yet to be done due to the preparation error problem, it may be allowed with a non-destructive 
measurement with a cavity.~\cite{PhysRevA.98.063619}}.
The current noise can be introduced as
\beq
S_p(\omega)&&=\int_{-\infty}^{\infty} d\tau e^{i\omega\tau}\langle \Delta I_p(\tau)\Delta I_p(0)+\Delta I_p(0)\Delta I_p(\tau) \rangle,
\nonumber\\
\eeq
where $\Delta I_p(\tau)=I_p(\tau)-\langle I_p(\tau) \rangle.$
At the level of the approximation used in this work, the current noise is expressed in terms of lesser and greater Green's functions as follows~\cite{PhysRevLett.82.4086}:
\beq
S_p(\omega)=\int d\tau e^{i\omega\tau}\Big[t_p^2G^<_{LR}(-\tau)G^>_{LR}(\tau) 
+\bar{t}_p^2G^<_{RL}(-\tau)G^>_{RL}(\tau) \nonumber\\
-|t_p|^2G^<_{RR}(-\tau)G^>_{LL}(\tau)-|t_p|^2G^<_{LL}(-\tau)G^>_{RR}(\tau) 
+(\tau\to-\tau)\Big].\nonumber\\
\eeq
We now focus on the white noise limit $\omega=0$, where a nontrivial relation between the current noise and DC current is expected.
Moreover, we can neglect the contribution of the normal tunneling of the NG modes  by focusing on the  low-temperature regime and
keeping only the contribution from the anomalous process.
Then, the current noise is reduced to the following simple expression:
\beq
S_p(0)\approx4{\cal T}^2G_{\text{an}}\Delta\mu,
\eeq
where we use $\coth(\beta\Delta\mu)\to1$.
Here we note that in the above expression,  a  contribution proportional to ${\cal T}^4$ is absent, which is correct up to $\Delta\mu$.
This is in sharp contrast to the shot noise of noninteracting particles
where ${\cal T}^4$ contributions bringing  bunching (anti-bunching) effects in bosons (fermions) is present.

\subsection{Absence of transport of the Nambu-Goldstone modes  in charged superconductors}
A fact that the current contributions  discussed in this work are negligible in charged superconductors can explicitly be shown as follows.
In the presence of the Coulomb interaction, the effective action of the NG modes is modified as~\cite{nagaosa2013} 
\beq
S_{\text{eff}}=\frac{\rho_s}{2\beta \Omega}\sum_kk^2\Big[\frac{\omega_n^2}{v^2k^2+\omega^2_p}+1\Big]\phi(-q)\phi(q),
\eeq
where $\displaystyle \sum_k\equiv\sum_{i\omega_n}\sum_{\mathbf{k}}$,  $\omega_n=2\pi n T$ with an integer $n$, and
\beq
\omega_p^2=\frac{4\pi e^2n}{M}
\eeq
is the plasma frequency.
By using the action above,  imaginary time Green's function of the NG mode is obtained as
\beq
{\cal G}(\omega_n)&&\equiv -4\int_0^{\beta}d\tau e^{i\omega_n\tau}\langle T_{\tau}[\phi(\tau)\phi(0)]\rangle_0\nonumber\\
&&=-\frac{4}{\rho_s\Omega}\sum_{\mathbf{k}}\frac{\omega_p^2+v^2k^2}{k^2(\omega_n^2+\omega_p^2+v^2k^2)}.
\eeq
As before, by using analytic continuation $i\omega_n\to\omega+i\eta$ with an infinitesimal positive parameter $\eta$, retarded Green's function is obtained as
\beq
&&g^R(\omega)=-\sum_{\mathbf{k}}\frac{2\sqrt{\omega_p^2+v^2k^2}}{\rho_s\Omega k^2}\times \nonumber\\
&&\Big[\frac{1}{\omega+\sqrt{\omega_p^2+v^2k^2}+i\eta}
-\frac{1}{\omega-\sqrt{\omega_p^2+v^2k^2}+i\eta}\Big].
\eeq
Therefore, its imaginary term is given by
\beq
&&\text{Im}[g^R(\omega)]=\sum_{\mathbf{k}}\frac{2\pi\sqrt{\omega_p^2+v^2k^2}}{\rho_s\Omega k^2}\times\nonumber\\
&&\Big[\delta(\omega+\sqrt{\omega_p^2+v^2k^2})
-\delta(\omega-\sqrt{\omega_p^2+v^2k^2})\Big].
\eeq
The expression above implies that the imaginary term vanishes unless $|\omega|\ge \omega_p$.
In contrast, the frequency $|\omega|\lessapprox \Delta\mu,T$ contributes to the mesoscopic current~\eqref{eq:general-mass}.
Since typically $\omega_p\gg \Delta>\Delta\mu,T$,
we conclude that the current contributions related to the NG modes are absent in charged superconductors.

\section{Summary}
By using the effective theory and tunneling Hamiltonian,
we have discussed DC transport of the NG modes in the fermionic superfluid point contact.
We have focused on the BCS regime and 
revealed the anomalous contribution in mass transport, which is the conversion process between the condensate and NG mode.
We also discussed that the anomalous contribution is not present in heat transport, which gives rise to breakdown of the Wiedemann-Franz law and the
absence of the bunching effect in current noise.

In addition to the BCS regime discussed in this work, it was recently shown that the anomalous contribution of the NG modes appears in the Bose-Einstein condensation regime~\cite{uchino}. Therefore, it is reasonable to expect that  this contribution
exists in the whole range of the BCS-Bose-Einstein condensation crossover.
 Indeed, the recent experiment on the current-biased Josephson junction implies the occurrence of a nonnegligible bias-independent  DC conductance at unitarity~\cite{Burchianti:2018aa,kwon2019strongly}.
At the same time, the recent ETH experiment observes a nonnegligible Seebeck coefficient~\cite{husmann2018}, which cannot be explained with the processes discussed
in this work. In order to address such a regime, therefore, a direct many-body calculation, which includes renormalization of Green's functions and vertex corrections rather than the effective field theory approach may be required.

\section*{acknowledgement}
The author thanks J.-P. Brantut and S. H\"{a}usler for discussions.
The author is supported by Matsuo Foundation, JSPS KAKENHI Grant Number JP17K14366 and a Waseda University
Grant for Special Research Projects (No. 2019C-461).

\appendix

\section{Correlation functions}
Here, we calculate the correlation functions in the bulk.
In the current calculation, following retarded Green's function is relevant:
\beq
g^R(\tau)=-i\theta(\tau)\langle [e^{2i\phi(\tau)},e^{-2i\phi(0)}]\rangle_0.
\eeq
As far as three dimensional configurations where phase fluctuations are small are concerned,
the correlation function above is approximated as
\beq
g^R(\tau)\approx -4i\theta(\tau)\langle [\phi(\tau),\phi(0)] \rangle_0.
\eeq
Similarly, the lesser Green's function that is another relevant quantity in the current calculation is obtained as
\beq
g^<(\tau)&&=-i\langle e^{-2i\phi(0)} e^{2i\phi(\tau)}\rangle_0\nonumber\\
&&\approx -i-4i\langle \phi(0)\phi(\tau)\rangle_0.
\eeq
Thus, up to this order of approximation, what we need to calculate is the correlation functions like  $\langle \phi(\pm\tau)\phi(0)\rangle_0$.
In order to evaluate these correlations, it is convenient to use the analytic continuation technique of imaginary time Green's functions.
To see this, we consider following  imaginary time Green's function:
\beq
{\cal G}(\tau)
&&=-4\langle T_{\tau}[\phi(\tau)\phi(0)]\rangle_0\nonumber\\
&&=-\frac{4}{(\beta\Omega)^2Z}\sum_{k_1,k_2}e^{-i\omega_n\tau}\int {\cal D}\phi \phi(k_1)\phi(k_2)
e^{-S_{\text{eff}}}\nonumber\\
&&=-\frac{4v^2}{\rho_s\beta\Omega}\sum_k\frac{e^{-i\omega_n\tau}}{\omega_n^2+v^2k^2},
\eeq
where $Z$ is the partition function,
$T_{\tau}$ is imaginary time-ordering operator~\cite{giamarchi},
and in the second line of equality we use the functional integrals in which time-ordered products are automatically ensured.
In the frequency space, the above correlation function is expressed as
\beq
{\cal G}(\omega_n)&&=\int_0^{\beta}d\tau e^{i\omega_n\tau}{\cal G}(\tau)\nonumber\\
&&=-\frac{4v^2}{\rho_s\Omega}\sum_{\mathbf{k}}\frac{1}{\omega_n^2+v^2k^2}.
\eeq
Then, retarded Green's function with real frequency $\omega$ can be obtained from analytic continuation $i\omega_n\to\omega+i\eta$ with
an infinitesimal positive parameter $\eta$. By doing this, we obtain
\beq
g^R(\omega)&&=-\frac{4v^2}{\rho_s\Omega}\sum_{\mathbf{k}}\frac{1}{2vk}\Big(\frac{1}{vk+\omega+i\eta}+\frac{1}{vk-\omega-i\eta}\Big)\nonumber\\
&&=-\frac{1}{v\pi^2\rho_s}\int_0^{v\Lambda}d(vk)\Big[\frac{vk}{vk+\omega+i\eta}+\frac{vk}{vk-\omega-i\eta}\Big],\nonumber\\
\eeq
where we introduce the momentum cutoff $\Lambda$  to restrict applicability of the tunneling Hamiltonian.
By using
\beq
\int dx\frac{x}{x\pm A}=x\mp A\log|x\pm A|,
\eeq
with a real constant $A$,
we reach
\beq
g^R(\omega)\approx-\frac{2\Lambda}{\pi^2\rho_s}-\frac{i\omega}{v\pi\rho_s},
\eeq
where we neglect $O(\omega^2)$ terms.
Similarly,  by using $i\omega_n\to\omega-i\eta$, advanced Green's function is obtained as
\beq
g^A(\omega)\approx-\frac{2\Lambda}{\pi^2\rho_s}
+\frac{i\omega}{v\pi\rho_s}.
\eeq

We finally calculate the lesser Green's function at real frequency,
\beq
g^<(\omega)
=-2\pi i\delta(\omega)-4i\int_{-\infty}^{\infty} d\tau e^{i\omega \tau}\langle \phi(0)\phi(\tau)\rangle_0.
\eeq
In order to obtain an useful expression of the second term in the right hand side, we consider the spectral representation of Green's function,
\beq
&&-4i\int_{-\infty}^{\infty} d\tau e^{i\omega \tau}\langle \phi(0)\phi(\tau)\rangle_0\nonumber\\
&&=-\frac{8\pi i}{Z}\sum_{n,m} e^{-\beta E_m}\langle n|\phi|m\rangle\langle m|\phi |n\rangle \delta(\omega+(E_n-E_m)),\nonumber\\
\eeq
where $\{| n\rangle\}$ or $\{| m\rangle\}$   is  the set of the energy eigenstates.  
Similarly, since
\beq
&&-4i\theta(\tau)\langle [\phi(\tau),\phi(0)]\rangle_0\nonumber\\
&&=-\frac{4i\theta(\tau)}{Z}\sum_{n,m}e^{i(E_n-E_m)\tau}\langle n|\phi |m\rangle
\langle m|\phi |n\rangle(e^{-\beta E_n}-e^{-\beta E_m}),\nonumber\\
\eeq
 the spectral representation of retarded Green's function is given by
\beq
g^{R}(\omega)=\frac{4}{Z}\sum_{n,m}\frac{\langle n|\phi |m\rangle
\langle m|\phi |n\rangle(e^{-\beta E_n}-e^{-\beta E_m})}{\omega+(E_n-E_m)+i\eta}.\nonumber\\
\eeq
Since its imaginary term is given by
\beq
\text{Im}[g^R(\omega)]&&=-\frac{4\pi}{Z}\sum_{n,m}e^{-\beta E_m} \langle n|\phi |m\rangle
\langle m|\phi |n\rangle(e^{\beta \omega}-1)\nonumber\\
&&\times\delta(\omega+(E_n-E_m)),
\eeq
we obtain
\beq
-4i\int_{-\infty}^{\infty} d\tau e^{i\omega \tau}\langle \phi(0)\phi(\tau)\rangle_0
&&=2i\text{Im}[g^R(\omega)]n(\omega).
\eeq
In total, the lesser Green's function is obtained as
\beq
g^<(\omega)=-2\pi i\delta(\omega)-\frac{2i\omega}{v\pi\rho_s}n(\omega).
\eeq
In a similar manner, greater Green's function is obtained as
\beq
g^>(\omega)=-2\pi i\delta(\omega)-\frac{2i\omega}{v\pi\rho_s}(1+n(\omega)).
\eeq

\bibliographystyle{apsrev4-1}
%

\end{document}